\documentclass[showpacs,preprintnumbers,amsmath,amssymb,prd,superscriptaddress,twocolumn,longbibliography]{revtex4-2}

\usepackage{dcolumn}
\usepackage{color}
\usepackage{subfigure}  
\def\comment#1{}

\usepackage{bm,latexsym,mathrsfs,enumerate,color}
\usepackage[mathcal]{euscript}
\usepackage[breaklinks=true,unicode=true,urlcolor = blue,colorlinks = true,citecolor = blue,linkcolor = blue]{hyperref}
\usepackage{graphicx}
\usepackage{todonotes}
\usepackage{wrapfig}
\usepackage{stmaryrd} 

\renewcommand{\vec}[1]{\bm{#1}}

\usepackage[breaklinks=true,unicode=true,urlcolor = blue,colorlinks = true,citecolor = blue,linkcolor = blue]{hyperref}

\DeclareMathAlphabet\mathbfcal{OMS}{cmsy}{b}{n}

\def\slashchar#1{\setbox0=\hbox{$#1$}           
	\dimen0=\wd0                                 
	\setbox1=\hbox{/} \dimen1=\wd1               
	\ifdim\dimen0>\dimen1                        
	\rlap{\hbox to \dimen0{\hfil/\hfil}}      
	#1                                        
	\else                                        
	\rlap{\hbox to \dimen1{\hfil$#1$\hfil}}   
	/                                         
	\fi}                                         %

\def\nablab{{\mbox{\boldmath $\nabla$}}}
\def\varphib{{\mbox{\boldmath $\varphi$}}}
\def\gammab{{\mbox{\boldmath $\gamma$}}}

\def\phib{{\mbox{\boldmath $\phi$}}}

\def\alphab{{\mbox{\boldmath $\alpha$}}}
\def\Pib{{\mbox{\boldmath $\Pi$}}}

\begin{document}

\title{Absence of induced magnetic monopoles in Maxwellian magnetoelectrics}
\author{Flavio S. Nogueira}
\affiliation{Institute for Theoretical Solid State Physics, IFW Dresden, Helmholtzstr. 20, 01069 Dresden, Germany}

\author{Jeroen van den Brink}
\affiliation{Institute for Theoretical Solid State Physics, IFW Dresden, Helmholtzstr. 20, 01069 Dresden, Germany}
\affiliation{Institut f\"ur Theoretische Physik and W\"urzburg-Dresden Cluster of Excellence ct.qmat, Technische Universit\"at Dresden, 01062 Dresden, Germany}

\date{Received \today}

\begin{abstract}
The electromagnetic response of topological insulators is governed by axion electrodynamics, which features a topological magnetoelectric term in the Maxwell equations. As a consequence magnetic fields become the source of electric fields and vice-versa, a phenomenon that is general for any material exhibiting a linear magnetoelectric effect. Axion electrodynamics has been associated with the possibility to create magnetic monopoles, in particular by an electrical charge that is screened above the surface of a magnetoelectric material. Here we explicitly solve for the electromagnetic fields in this geometry and show that while vortex-like magnetic screening fields are generated by the electrical charge their divergence is identically zero at every point in space which implies an absence of induced magnetic monopoles. Nevertheless magnetic image charges can be made explicit in the problem and even if no bound state with electric charges yielding a dyon arises, a dyon-like angular momentum follows from our analysis. Because of its dependence on the dielectric constant this angular momentum is not quantized, which is consistent with a general argument that precludes magnetic monopoles to be generated in Maxwell magnetoelectrics.  
We also solve for topologically protected zero modes in the Dirac equation  
induced by the point charge. Since the induced topological defect on the TI surface carries an electric charge as a result of the axion term, 
these zero modes are not self-conjugated.
\end{abstract}

\maketitle

\section{Introduction}

A remarkable feature of three-dimensional topological insulators (TIs) is their so-called magnetoelectric (ME) effect, a collection of phenomena where magnetic fields become the source of electric fields and vice-versa \cite{Qi-2008}. This topological electromagnetic response is governed by so-called axion electrodynamics, which features a magnetoelectric term ${\cal L}_a=\alpha\theta/(4\pi^2)\vec{E}\cdot\vec{B}$ in  the Lagrangian density ${\cal L}_a$, with electric and magnetic fields $\vec{E}$ and $\vec{B}$ respectively, $\theta$  a $2\pi$-periodic parameter and $\alpha$ the fine-structure constant.  In a topological insulator $\theta$ is a parameter that follows from the band structure topology, being given by a Berry non-Abelian flux in the Brillouin zone \cite{Qi-2008,Essin_2009}. By symmetry the magnetoelectric coupling term is actually present in any material that exhibits a linear magnetoelectric effect -- induction of magnetization by an electric field or of electric polarization by a magnetic field. However, in ordinary magnetoelectric materials such as Cr$_2$O$_3$, BiFeO$_3$, and GdAlO$_3$ the magnetoelectric coupling constants are quite small~\cite{Coh2011}. The topological ME effect has been recently measured using Faraday and Kerr rotation \cite{Axion-exp-Armitage,Axion-exp-Molenkamp},  which were shown to be quantized according to the prediction of axion electrodynamics of TIs. 

A number of further interesting consequences of the axion term have been predicted, for instance that a cylindrical TI becomes electrically polarized under an applied magnetic field parallel to the cylinder symmetry axis \cite{Ryu-axion}. An interesting possible experimental setup exploring this effect is a flux tube piercing the interior of a TI, perpendicular to its surfaces \cite{Franz-solenoid}. If the surfaces are coated with thin film ferromagnets with opposite magnetizations, the surface states become gapped and a topological electromagnetic response ensuing the axion term in the Lagrangian occurs.  In this scenario the cylinder becomes an Aharonov-Bohm flux tube and an electrical polarization is induced leading to fractional charges $\pm e/2$ on the top and bottom surfaces, respectively \cite{Franz-solenoid}.  
In the case of a magnetic vortex that enters from a superconductor (SC) into a  time-reversal invariant TI, it was shown that the vortex induces a charge of $ e/4$  \cite{Nogueira-Nussinov-van_den_Brink-PRL,Nogueira-van_den_Brink-Nussinov_London_axion} at the SC-TI interface. In this situation also the vortex angular momentum, which determines the vortex statistics, is fractional \cite{Nogueira-van_den_Brink-Nussinov_London_axion}. The emergence of fractional charges is reminiscent of the Witten effect \cite{Witten}, which predicts that the axion term causes electric charge fractionalization in the presence of magnetic monopoles. When real magnetic monopoles were to be present in an axion magnetoelectric, fractional electric charges would occur not only at surfaces but also in the bulk of a magnetoelectric, since in this case the Maxwell equations are modified despite the axion Lagrangian ${\cal L}_a$ being a total derivative \cite{Wilczek}.

\begin{figure}
 	\includegraphics[width=.7\columnwidth]{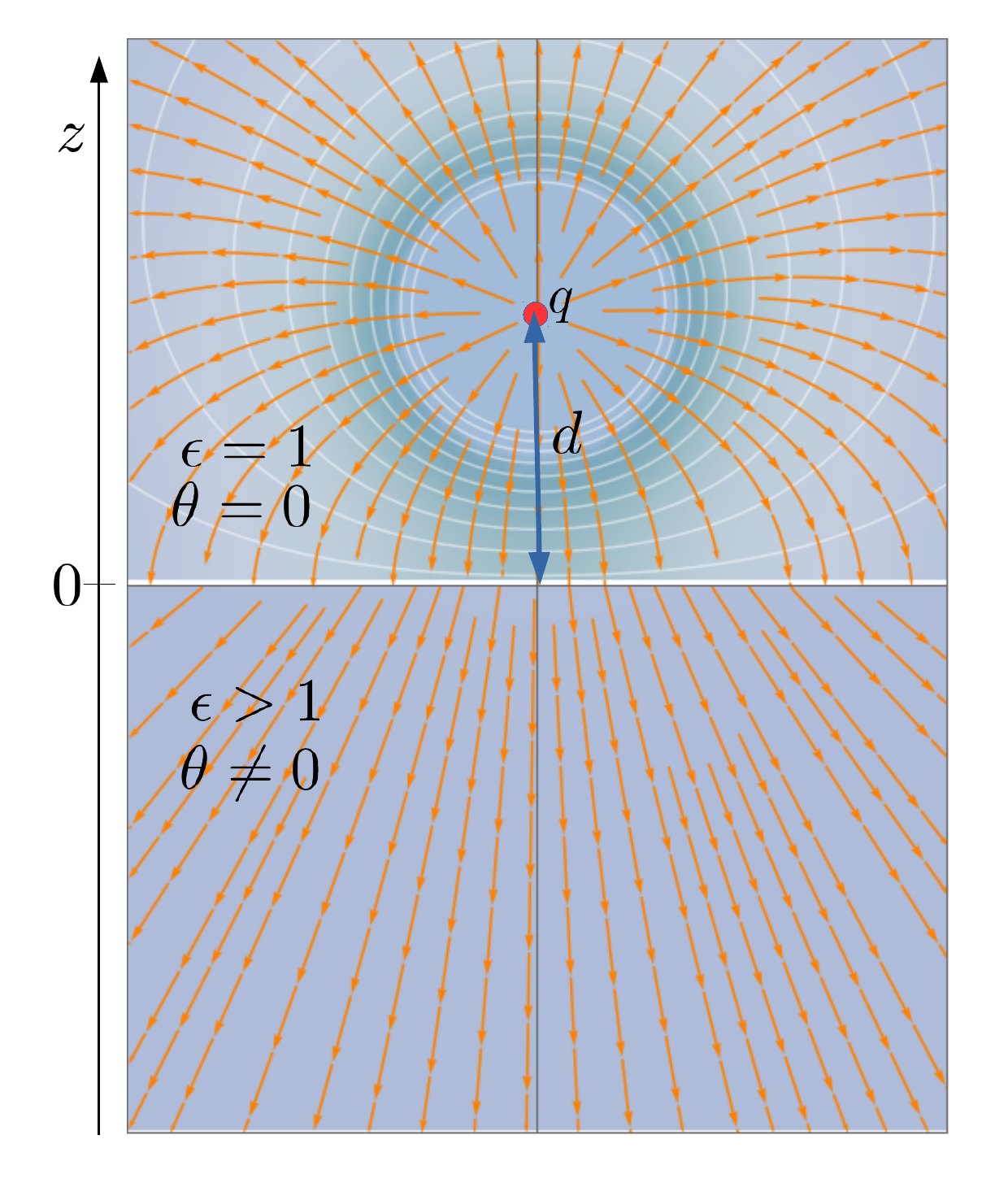}
\caption{Point charge $q$ at a distance $d$ above the surface of a (topological) magnetoelectric with $\theta \neq 0$ and dielectric constant $\epsilon>1$ occupying the region $z<0$. The region  $z>0$ is (topologically trivial) vacuum. The electric field lines are shown.}
\label{Fig:Pointcharge}
 \end{figure}

In this context it is highly interesting that the presence of  a magnetoelectric term in the Maxwell Lagrangian has been associated with the possibility 
of creating magnetic monopoles. In particular, the situation has been considered in which a magnetic monopole emerges from the screening of an electrical charge that is situated outside a magnetoelectric material, at a certain distance $d$ from its the surface~\cite{Qi_image_monopole,Karch}, see Fig.~\ref{Fig:Pointcharge}. This would be quite remarkable as it would imply that the condition that the magnetic field be divergence-free is lifted by the axion term in one way or another.  
Here we revisit this very well defined geometry and determine the unique solution for the electric and magnetic fields by direct evaluation, without resorting to an image charge construction~\cite{Qi_image_monopole} or Green function formalism~ \cite{Ruiz-PhysRevD.94.085019} and determine also the angular momentum of this dyon-like object carrying both electric and magnetic charge. The resulting divergence of the magnetic field vanished at every point in space. Induced magnetic monopoles are thus absent and instead the electrical charge generates a magnetic vortex structure near the magnetoelectric surface which in turn generates magnetic screening fields in all of space. In the limit that the electrical charge is placed at the TI surface ($d \rightarrow 0$) it almost behaves like a magnetic monopole, but actually corresponds to a point vortex, or Pearl vortex \cite{Pearl-1964}, still satisfying the local constraint $\nablab\cdot\vec{B}=0$.  Although the solution can be cast in terms of image electric and magnetic charges, the latter are not related by a Dirac duality quantization characteristic of magnetic monopoles and hence cannot be interpreted as such. We will show that despite this difficulty dyon-like behavior occurs. Dyons are dipoles constituted of an electric and a magnetic charge and are well known to have an angular momentum \cite{Schwinger-1969} with the following properties: (i) its value is independent of the separation between the electric and magnetic charges, (II) it is nonzero even if the dyon is at rest, and (iii) it is quantized of the  Dirac duality relation $eg/c=n\hbar/2$, $n\in\mathbb{Z}$. For our problem of a point charge a distance $d$ apart from the TI surface, the resulting angular momentum fulfils properties (i) and (ii), but not (iii), which is signalled by the angular momentum explicitly depending on the dielectric constant $\epsilon$ of the TI. Insisting the angular momentum be quantized, would imply the (static) dielectric constant be negative, violating the inequality $\epsilon>1$.  This dependence on $\epsilon$ prevents the interpretation of this dyon-like object as an anyon, a quasi-particle having fractional statistics as quantum statistics is a universal property of (quasi-)particles and should not depend on the details of the medium they are embedded in.   

The outline of the paper is as follows.
We first introduce in Section ~\ref{Sec:AMequations} the Axion Maxwell equations for a semi-infinite magnetoelectric. 
In order to better understand the nature of the actual topological defect being induced by the point charge on the TI
Section \ref{Sec:Vortex} we derive an exact solution for a vortex of finite length $L$ in the London limit and obtain the 
Pearl vortex solution in the limit $L\to 0$ and large planar distances compared to the London 
penetration depth. Such a Pearl vortex regime will arise as a topological defect 
induced by a point charge sitting precisely on the TI surface. This result will be obtained as a special case of the more general solution 
in Section \ref{Sec:Solution}. In Section \ref{Subsec:solenoid} we compare the solution of a very thin solenoid of finite length $L$ to the obtained solution for 
a vortex of finite length and notice crucial differences. At large distances the field profiles arising from the end points of the solenoid behave 
precisely as magnetic monopoles. In the vortex case the monopole-like field profiles do not quite resemble actual monopoles due to Meissner 
screening inside the superconductor. Such a screening is absent in the case of a solenoid.
The general solution of the electric and magnetic field and the associate angular momentum for a charge placed at a distance $d$ away from such a  semi-infinite magnetoelectric medium is detailed in Section~\ref{Sec:Solution}. In Section \ref{Sec:zero-modes} we derive the zero energy mode solutions 
of the Dirac equation on the TI surface in the presence of the electromagnetic field  
induced by the external point charge.
A general argument for absence of magnetic monopoles is provided  in~\ref{Sec:General} and we end with a brief summary and conclusions.

\section{Axion Maxwell equations for a semi-infinite magnetoelectric}
\label{Sec:AMequations}
Given the textbook nature of the problem on one the hand and the importance of its exact solution on the other, we present the steps to obtain a direct solution for the electric and magnetic fields in a semi-infinite three-dimensional Maxwellian magnetoelectric in some detail. The effective Lagrangian density is given by \cite{Qi-2008},
\begin{equation}
{\cal L}=\frac{1}{8\pi}\left(\epsilon\vec{E}^2-\frac{1}{\mu}\vec{B}^2\right)-{\cal L}_a,
\end{equation}
where ${\cal L}_a$ is given above and Gaussian units are being used. Similarly to Ref. \cite{Qi_image_monopole}, we assume that the (toplogical) magnetoelectric medium occupies the region $z\leq 0$, with the surface at $z=0$ separating it from a trivial insulator, which we assume to be the vacuum, see Fig.~\ref{Fig:Pointcharge}. Thus, we have a dielectric constant $\epsilon=1$ and $\theta=0$ for $z>0$.  We further assume for simplicity that the magnetic properties are such that $\mu=1$ for all $z$. 
The easiest way to obtain the 
field equations is to write the standard Maxwell equations in the presence of matter and recall that, 
\begin{equation}
\label{Eq.Cosnt-eqs}
\vec{D}=4\pi\frac{\partial{\cal L}}{\partial\vec{E}}, ~~~~~~~~~\vec{H}=-4\pi\frac{\partial{\cal L}}{\partial \vec{B}}.
\end{equation}
We obtain in this way the general field equations in the form,  
%
\begin{eqnarray}
\label{Eq:Gauss}
\nablab\cdot\left(\epsilon\vec{E}-\frac{\alpha\theta}{\pi}\vec{B}\right)&=&4\pi\rho, \\
%
\label{Eq:Ampere}
\nablab\times\left(\vec{B}+\frac{\alpha\theta}{\pi}\vec{E}\right)&=&\frac{4\pi}{c}{\bf j}+\frac{1}{c}\partial_t\left(\epsilon\vec{E}-\frac{\alpha\theta}{\pi}\vec{B}\right),
\end{eqnarray}
while the source-free Maxwell equations remain unchanged,  
\begin{equation}
\nablab\times\vec{E}=-\frac{1}{c}\partial_t\vec{B},~~~~~~~~~~~\nablab\cdot\vec{B}=0,
\end{equation}
since the latter are actually constraints following from the Bianchi identity for the electromagnetic field strength. 
 Die to the second Eq. (\ref{Eq.Cosnt-eqs}) above, $\vec{H}=\vec{B}+(\alpha\theta/\pi)\vec{E}$, we see that one way of solving the 
 problem shown in Fig. \ref{Fig:Pointcharge} is to use a magnetic image charge \cite{Qi_image_monopole}, since the problem 
 becomes analogous to the one of a semi-infinite system with a magnetic permeability $\mu\neq 1$ \cite{zangwill2013modern}. 
 Indeed, here the role of magnetization is played by $-\alpha\theta\vec{E}/(4\pi^2)$, so we can use the constraint 
 $\nablab\cdot\vec{B}=0$ to define a magnetic charge density, $\rho_M=\nablab\cdot [\alpha\theta\vec{E}/(4\pi^2)]$, leading to 
 the equations $\nablab\cdot\vec{H}=4\pi\rho_M$ and $\nablab\times\vec{H}=0$, which is formally identical to a problem 
 in electrostatics \cite{zangwill2013modern,neumann1883hydrodynamische}. We will not follow this approach here and proceed 
 to solve the equations using a vector potential, as this will help us to clarify the similarities and differences from these image 
 magnetic charges and actual magnetic monopoles. 
  
 After setting $\vec{E}=-\nablab\phi$ and $\vec{B}=\nablab\times\vec{A}$, we obtain the differential equations for the scalar and vector 
 potentials, 
 %
  \begin{eqnarray}
 \label{Eq:Poisson-1}
 -\nabla^2\phi-\frac{\alpha}{\pi}\nablab\theta\cdot(\nablab\times\vec{A})&=&4\pi\rho\quad\quad (z>0), \\
 %
 \label{Eq:Poisson-2}
 -\epsilon\nabla^2\phi-\frac{\alpha}{\pi}\nablab\theta\cdot(\nablab\times\vec{A})&=&4\pi\rho\quad\quad (z<0),\\
 %
 -\nabla^2\vec{A}-\frac{\alpha}{\pi}\nablab\theta\times\nablab\phi&=&0,
 \end{eqnarray}
 where we have assumed the Coulomb gauge $\nablab\cdot\vec{A}=0$. Since $\nablab\theta=-\theta\delta(z)\hat{\vec{z}}$ for the system under 
 consideration, we have to actually solve the equations, 
 \begin{equation}
 \label{Eq:Diff-potentials}
 -\nablab\cdot(\epsilon\nablab\phi)=4\pi\rho,~~~~~~~~~~~~~~~~\nabla^2\vec{A}=0, 
 \end{equation}
 subjected to boundary conditions reflecting the discontinuities in the normal derivatives of the potentials at $z=0$, 
 implied also by  the change of 
 $\theta$ at the interface.  
 Translational invariance in the $xy$-plane implies, 
 \begin{equation}
 -\frac{d^2\hat{\vec{A}}}{dz^2}+\vec{p}^2\hat{\vec{A}}(\vec{p},z)=0,
 \end{equation}
 where $\hat{\vec{A}}(\vec{p},z)$ is the Fourier transform of the vector potential in the plane. The above equation is to be solved with the 
 boundary conditions, 
\begin{eqnarray}
 \hat{\vec{A}}(\vec{p},+\eta)&=&\hat{\vec{A}}(\vec{p},-\eta), \\
 %
 \left.\frac{d\hat{\vec{A}}}{dz}\right|_{z=-\eta}-\left.\frac{d\hat{\vec{A}}}{dz}\right|_{z=+\eta}&=&\frac{\alpha\theta}{\pi}
 (\hat{\vec{z}}\times\hat{\vec{E}}(\vec{p},z=0)),
 \end{eqnarray}
 where $\eta\to 0+$. One finds after a straightforward calculation that
 \begin{equation}
 \label{Eq:A}
 \vec{A}(\vec{r},z)=\frac{\alpha\theta}{4\pi^2}\int d^2r'\frac{\hat{\vec{z}}\times\vec{E}(\vec{r}',z'=0)}{\sqrt{(\vec{r}-\vec{r}')^2+z^2}},
 \end{equation}
which yields 
 \begin{eqnarray}
 \label{Eq:B}
 &&\vec{B}(\vec{r},z)=\nablab\times\vec{A}=\frac{\alpha\theta}{4\pi^2}\left\{z\int d^2r'\frac{\vec{E}(\vec{r}',z'=0)}{[(\vec{r}-\vec{r}')^2+z^2]^{3/2}}
 \right.\nonumber\\
 &&-\left.\hat{\vec{z}}\int d^2r'\frac{(\vec{r}-\vec{r}')+z\hat{\vec{z}}}{[(\vec{r}-\vec{r}')^2+z^2]^{3/2}}\cdot \vec{E}(\vec{r}',z'=0)	\right\}.
 \end{eqnarray}
 At this point it is important to observe that the magnetic field is divergence-free everywhere in space as the above expression obviously satisfies $\nablab\cdot \vec{B}=0$ everywhere, irrespective of the form of the electric field.  
%
 %

 \section{Vortex of finite length, solenoids and artificial monopoles}
 \label{Sec:Vortex}
 
 In order to put better in perspective what kind of solution is actually obtained and understand how objects that from large distance 
 may look like a magnetic monopole, we first consider a vortex line of finite length in the London limit. In the limit where the length of the vortex 
 line approaches zero, one obtains the field of a so called Pearl vortex at large planar distances, $r\gg \lambda_L$, where $\lambda_L$ is the 
 London penetration depth. We will see later that for a TI a Pearl vortex-like solution \cite{Pearl-1964} is obtained for the case where the point charge sits precisely 
 at the TI surface. The major difference to the actual Pearl vortex in a thin superconducting slab is that the solution for the TI will be exact and 
 not only valid at large distances in the plane. 
 
 \subsection{Brief review of London theory}
 
 In the London limit the static superconducting current is given by the well known formula \cite{tinkham2004introduction}, 
 \begin{equation}
 \vec{j}_s(\vec{r},z)=2e\rho_s\vec{v}_s(\vec{r},z),
 \end{equation}
 where $\rho_s$ is the superfluid density and $\vec{v}_s$ is the superfluid velocity, which in the case of a superconductor is given by, 
 \begin{equation}
 \vec{v}_s(\vec{r},z)=\frac{1}{m}\left[\hbar\nablab\varphi-\frac{2e}{c}\vec{A}(\vec{r},z)\right],
 \end{equation}
 where $\varphi$ is the phase of the superconducting order field. For later use, 
 we are labeling the planar coordinates $\vec{r}=(x,y)$ separately from $z$. Thus, the London equation is simply the Maxwell equation, 
 \begin{equation}
 \nablab\times\vec{B}=\frac{4\pi}{c}\vec{j}_s,
 \end{equation} 
 supplemented by the constraint, $\nablab\cdot\vec{B}=0$. 
 
 If the space is simply connected, we have, $\nablab\times\nablab\varphi=0$ everywhere, and the London equation simplifies to, 
 \begin{equation}
 -\nabla^2\vec{B}+m_L^2\vec{B}=0,
 \end{equation}
 where, 
 \begin{equation}
 m_L^2=\frac{16\pi e^2\rho_s}{mc^2}.
 \end{equation}
 The latter equation yields the London penetration depth $\lambda_L=m_L^{-1}$. Vortices make space multiply connected and the London equation 
 becomes more interesting. Generally a superconductor features both open and closed (loops) vortex lines \cite{kleinert1989gauge}. If vortex lines are 
 accounted for, the phase gradient has the form \cite{Fetter_PhysRev.162.143},
 \begin{equation}
 \label{Eq:Phase-grad}
 \nablab\varphi(\vec{R})=\nablab\varphi_L+\frac{1}{2}\sum_{i=1}^{N_v}\int_{\mathcal{L}_i}\frac{d\gammab_i\times(\vec{R}-\gammab_i)}{|\vec{R}-\gammab_i|^3},
 \end{equation}
 where $\vec{R}=(\vec{r},z)$. 
 In the above equation $\varphi_L$ denotes the longitudinal part of the phase satisfying $\nablab\times\nablab\varphi_L=0$, while the 
 second term corresponds to the contribution from vortex lines $\mathcal{L}_i$, $1\leq i\leq N_v$, with the integral being along the $i$-th vortex line 
 determined by the vector $\gammab_i$. Due to the second term, we have $\nablab\times\nablab\varphi\neq 0$. For the simple case of a single infinite 
 straight  vortex line, we have $\gammab(z')=z'\hat{\vec{z}}$, $z'\in(-\infty,\infty)$, such that we obtain, 
 \begin{equation}
 \nablab\varphi=\nablab\varphi_L+\frac{\hat{\vec{z}}\times\vec{r}}{r^2}.
 \end{equation}
 Thus, 
 \begin{equation}
 \oint_C d\vec{R}\cdot\nablab\varphi=\oint_C d\vec{r}\cdot\frac{\hat{\phib}}{r}=\int_{0}^{2\pi n}d\phi=2\pi n, 
 \end{equation} 
 where $n\in\mathbb{Z}$ is the vorticity winding number. From Stokes theorem we can therefore write, 
 \begin{equation}
 \nablab\times\nablab\varphi=2\pi n\delta^2(\vec{r})\hat{\vec{z}},
 \end{equation} 
 and the London equation becomes \cite{tinkham2004introduction}, 
 \begin{equation}
 -\nabla^2\vec{B}+m_L^2\vec{B}=m_L^2n\Phi_0\delta^2(\vec{r})\hat{\vec{z}},
 \end{equation}
 where $\Phi_0=hc/(2e)$ is the elementary flux quantum for a superconductor, corresponding to half of the usual flux quantum arising in the 
 Aharonov-Bohm effect. 
 Therefore, for an infinite system the ANO solution \cite{A-vortices,NO-vortices} for a single infinite vortex line is given in the London limit by \cite{tinkham2004introduction}, 
 \begin{equation}
 \label{Eq:B-vortex}
 \vec{B}(\vec{r})=\frac{n\Phi_0}{2\pi}m_L^2K_0(m_Lr)\hat{\vec{z}},
 \end{equation}
 where $K_0(x)$ is a modified Bessel function of second kind. 
 
 \subsection{London theory in a superconducting slab}
 
 For a superconducting slab of thickness $L$ general vortex line solutions in the London regime have been obtained by Brandt \cite{brandt1981properties} 
 and Carneiro and Brandt 
 \cite{Carneiro-Brandt_PhysRevB.61.6370}. For a straight vortex line parallel to the $z$-axis we have that vector potential 
 now depends on $z$ and has the form, 
 \begin{equation}
 \vec{A}(\vec{r},z)=A(r,z)\frac{\hat{\vec{z}}\times\vec{r}}{r}.
 \end{equation}
 We assume that the slab occupies the region $\mathcal{R}=\{(\vec{r},z)\in\mathbb{R}^3~|~\vec{r}\in\mathbb{R}^2\wedge z\in[-L,0]\}$. Thus, 
 the London equation has to be solved in cylindrical coordinates using boundary conditions for $A(r,z)$ both at $z=0$ and at $z=-L$. These are 
 continuity of $A(r,z)$ of its derivative with respect to $z$ at the surfaces $z=0$ and $z=-L$. The region outside the slab is assumed to be vacuum. 
 The easiest way to obtain the solution is to recall the vector potential leading to the magnetic field (\ref{Eq:B-vortex}) and generalize it to, 
 \begin{equation}
 \label{Eq:A}
 A(r,z)=\frac{n\Phi_0m_L^2}{2\pi}\int_{0}^{\infty}dp\frac{J_1(pr)a(p,z)}{p^2+m_L^2},
 \end{equation}
 where $J_1(x)$ is a Bessel function and 
 $a(p,z)$ is determined by the boundary conditions. The infinite system has a solution corresponding to $a(p,z)=1$. In this case the integral can be performed 
 exactly to obtain, 
 \begin{equation}
 A(r)=\frac{n\Phi_0}{2\pi}\left[\frac{1}{r}-m_LK_1(m_Lr)\right],
 \end{equation}
 which yields the vector potential whose curl produces the magnetic field (\ref{Eq:B-vortex}). 
 
 For a system of thickness $L$ we obtain on the other hand,
 \begin{equation}
 a(p,z)=
 \begin{cases}
 \frac{\epsilon(p)e^{-pz}}{\epsilon(p)+p\coth[L\epsilon(p)/2]}, & z>0\\[10pt]
 1-\frac{p\cosh[\epsilon(p)(z+L/2)]}{p\cosh[L\epsilon(p)/2]+\epsilon(p)\sinh[L\epsilon(p)/2]}, & -L<z<0\\[10pt]
 \frac{\epsilon(p)e^{p(z+L)}}{\epsilon(p)+p\coth[L\epsilon(p)/2]}, & z<-L
 \end{cases}
 \end{equation}
 where $\epsilon(p)=\sqrt{p^2+m_L^2}$. In this case the integral in Eq. (\ref{Eq:A}) cannot be performed in closed form. 
 
 A regime particularly interesting for us is the large distance one, where $r\gg \lambda_L$. In this regime we obtain, 
 \begin{equation}
 A(r,z)\approx\frac{n\Phi_0}{2\pi r}
 \begin{cases}
 1-\frac{z}{\sqrt{r^2+z^2}}, & z>0\\[10pt]
 1, & -L<z<0\\[10pt]
 1+\frac{z+L}{\sqrt{r^2+(z+L)^2}}, & z<-L
 \end{cases}
 \end{equation}
 and thus, 
 \begin{equation}
 \label{Eq:B-vortex-large-dist}
 \vec{B}(\vec{R})\approx\frac{n\Phi_0}{2\pi}
 \begin{cases}
 \frac{\vec{R}}{R^3}, & z>0\\[10pt]
 0, & -L<z<0\\[10pt]
 -\frac{(\vec{R}+L\hat{\vec{z}})}{|\vec{R}+L\hat{\vec{z}}|^3}, & z<-L
 \end{cases}
 \end{equation}
 corresponding to the magnetic field of a magnetic monopole of charge $g=n\Phi_0/(2\pi)$ located at $\vec{R}_0=0$ for $z>0$ and 
 to the magnetic field of a magnetic monopole of charge $-g$ located at $\vec{R}_1=(0,0,-L)$ for $z<-L$. If in addition we 
 consider  the thin film limit $L\to 0$ we obtain a point vortex field profile, a so called Pearl vortex \cite{Pearl-1964},
 \begin{equation}
 \label{Eq:Pearl-vortex}
 \vec{B}(\vec{R})\approx\frac{n\Phi_0}{2\pi}{\rm sgn}(z)\frac{\vec{R}}{R^3}. 
 \end{equation}
 
 Note that despite exhibiting for both $z>0$ and $z<-L$ fields of magnetic monopoles, the interval restrictions guarantee that $\nablab\cdot\vec{B}=0$, 
 as it should. 
 
 Another limit case of interest is the short distance one corresponding to $r\ll\lambda_L$. In this case, 
 \begin{equation}
 A(r,z)\approx\frac{n\Phi_0r}{4\pi \lambda_L^2}
 \begin{cases}
 1-\frac{z}{\sqrt{r^2+z^2}}, & z>0\\[10pt]
 1, & -L<z<0\\[10pt]
 1+\frac{z+L}{\sqrt{r^2+(z+L)^2}}, & z<-L
 \end{cases}
 \end{equation}
 which leads to the typical thin solenoid expression for the magnetic field in the slab region $\mathcal{R}$, 
 \begin{equation}
 \vec{B}(\vec{R})=\frac{n\Phi_0}{2\pi\lambda_L^2}\hat{\vec{z}}. 
 \end{equation}
 
 \subsection{Comparison with a thin solenoid of length $L$}
 \label{Subsec:solenoid}
 
 The asymptotic behavior of the vortex solution for a slab geometry exhibited some similarities with a solenoid. There are some crucial differences, however, 
 which we analyze below. 
 
 It is a well known fact in classical electrodynamics that a  thin solenoid can be thought as a line of point magnetic dipoles, 
 an approximation valid at large distances \cite{zangwill2013modern}. 
 The vector potential is assumed to be a sum over infinitesimal elements of vector potential associated to an element of magnetic moment, 
 $d\vec{m}=m_0dz'\hat{\vec{z}}$, where $m_0$ is the magnetic moment per unit length of the solenoid, assumed to lie along the $z$-axis.  Thus, 
 \begin{equation}
 \label{Eq:dA}
 d\vec{A}=\frac{d\vec{m}\times(\vec{R}-z'\hat{\vec{z}})}{|\vec{R}-z'\hat{\vec{z}}|^3}.
 \end{equation} 
 The vector potential for a thin solenoid of length $L$ is therefore given by integrating the above equation in $z'\in(-L,0)$, 
 \begin{eqnarray}
 \vec{A}(\vec{r},z)&=&m_0r\hat{\phib}\int_{-L}^{0}\frac{dz'}{[r^2+(z-z')^2]^{3/2}}
 \nonumber\\
 &=&\frac{m_0\hat{\phib}}{r}\left[\frac{z+L}{\sqrt{r^2+(z+L)^2}}-\frac{z}{\sqrt{r^2+z^2}}\right].
 \end{eqnarray}
 This yields the magnetic field, 
 \begin{eqnarray}
 \label{Eq:B-solenoid}
 \vec{B}(\vec{R})&=&m_0\left[\frac{\vec{R}}{R^3}-\frac{(\vec{R}+L\hat{\vec{z}})}{|\vec{R}+L\hat{\vec{z}}|^3}\right]
 \nonumber\\
 &+&4\pi m_0\delta^2(\vec{r})[\theta_H(-z)-\theta_H(z+L)]\hat{\vec{z}},
 \end{eqnarray}
 where $\theta_H(x)$ is the Heaviside unit step function. Thus, a thin solenoid of length $L$ at large distances appears as two magnetic 
 monopoles of charges $\pm m_0$ connected by a Dirac string of length $L$. Note that the presence of the string guarantees that 
 $\nablab\cdot\vec{B}=0$ 
 \footnote{Compare with the discussion for a semi-infinite solenoid at page 344 of Zangwill's book \cite{zangwill2013modern} and exercise 11.5 at page 351.}.  
 
 The magnetic field (\ref{Eq:B-solenoid}) is clearly very different from the magnetic field of a vortex line at large distances, Eq. (\ref{Eq:B-vortex-large-dist}). 
 In the latter equation the magnetic fields of the monopoles are completely screened inside the slab. This is not the case in Eq. (\ref{Eq:B-solenoid}). Also the 
 limit $L\to 0$ does not yield the magnetic field of a Pearl vortex, yielding instead, 
 \begin{equation}
 \vec{B}(\vec{R})|_{L=0}=-4\pi m_0{\rm sgn}(z)\delta^2(\vec{r})\hat{\vec{z}}. 
 \end{equation}
 
 Interestingly, the result for a straight thin solenoid can be generalized to any curved shape. This can be done by exploring the similarity of Eq. (\ref{Eq:dA}) with 
 Eq. (\ref{Eq:Phase-grad}) for the phase gradient in a superfluid. Accordingly, we consider a line of infinitesimal magnetic moments along a curved line 
 $\mathcal{C}$ defined by the vector function, 
 $\gammab(s)$, $s\in[0,1]$, i.e., $d\vec{m}=m_0d\gammab$.  Thus, instead of Eq. (\ref{Eq:dA}) we have, 
 \begin{equation}
 \label{Eq:dA-curved}
 d\vec{A}=m_0\frac{d\gammab\times(\vec{R}-\gammab)}{|\vec{R}-\gammab|^3},
 \end{equation} 
 so that, 
 \begin{equation}
 \vec{A}(\vec{R})=m_0\int_{\mathcal{C}}\frac{d\gammab\times(\vec{R}-\gammab)}{|\vec{R}-\gammab|^3}. 
 \end{equation}
 The magnetic field is therefore derived as follows, 
 \begin{eqnarray}
 \epsilon_{ijk}\partial_jA_k(\vec{R})&=&m_0\epsilon_{ijk}\partial_j\int_{\mathcal{C}}\frac{\epsilon_{klm}d\gamma_l(x_m-\gamma_m)}{|\vec{R}-\gammab|^3}
 \nonumber\\
 &=&m_0(\delta_{il}\delta_{jm}-\delta_{im}\delta_{jl})\partial_j\int_{\mathcal{C}}\frac{d\gamma_l(x_m-\gamma_m)}{|\vec{R}-\gammab|^3}
 \nonumber\\
 &=&-m_0\int_{\mathcal{C}}d\gamma_j\frac{\partial}{\partial x_j}\frac{(x_i-\gamma_i)}{|\vec{R}-\gammab|^3}
 \nonumber\\
 &+&m_0\int_{\mathcal{C}} d\gamma_i\underbrace{\nablab\cdot\frac{\vec{R}-\gammab}{|\vec{R}-\gammab|^3}}_{=4\pi\delta^3(\vec{R}-\gammab)}
 \nonumber\\
 &=&m_0\int_{0}^1ds\frac{d\gamma_j}{ds}\frac{\partial}{\partial\gamma_j}\frac{(x_i-\gamma_i)}{|\vec{R}-\gammab|^3}
 \nonumber\\
 &+&4\pi m_0\int_{\mathcal{C}}d\gamma_i\delta^3(\vec{R}-\gammab)
 \nonumber\\
 &=&m_0\int_{0}^{1}ds\frac{d}{ds}\frac{(x_i-\gamma_i(s))}{|\vec{R}-\gammab(s)|^3}
 \nonumber\\
 &+&4\pi m_0\int_{\mathcal{C}}d\gamma_i\delta^3(\vec{R}-\gammab),
 \end{eqnarray}
 which immediately leads to, 
 \begin{eqnarray}
 \label{Eq:B-solenoid-curved}
 \vec{B}(\vec{R})&=&m_0\left[\frac{(\vec{R}-\gammab_2)}{|\vec{R}-\gammab_2|^3}-\frac{(\vec{R}-\gammab_1)}{|\vec{R}-\gammab_1|^3}\right]
 \nonumber\\
 &+&4\pi m_0\int_{\mathcal{C}}d\gammab\delta^3(\vec{R}-\gammab),
 \end{eqnarray}
 where $\gammab_2=\gammab(1)$ and $\gammab_1=\gammab(0)$. Equation (\ref{Eq:B-solenoid-curved}) obviously includes Eq. (\ref{Eq:B-solenoid}) 
 as a special case. Indeed, in this case we have simply $\gammab(z')=z'\hat{\vec{z}}$, with $\gammab_2=0$ and $\gammab_1=-L\hat{\vec{z}}$, while for 
 the delta function term we have, 
 \begin{eqnarray}
 &&\int_{\mathcal{C}}d\gammab\delta^3(\vec{R}-\gammab)=\hat{\vec{z}}\delta^2(\vec{r})\int_{-L}^0dz'\delta(z-z')
 \nonumber\\
 &=&\hat{\vec{z}}\delta^2(\vec{r})\int_{-\infty}^\infty dz'[\theta_H(-z')-\theta_H(z'+L)]\delta(z-z')
 \nonumber\\
 &=&\delta^2(\vec{r})[\theta_H(-z)-\theta_H(z+L)]\hat{\vec{z}}. 
 \end{eqnarray}

 \section{Solution for an electric point charge above a magnetoelectric}
 \label{Sec:Solution}
 \subsection{Calculation of the electric field}
We now consider for the charge density a point charge $q$ at $z=d>0$ as indicated in Fig.\ref{Fig:Pointcharge}. 
 Since the Poisson equation is translation invariant in the $xy$-plane, one can perform a two-dimensional Fourier transform to obtain the differential equations for the Fourier-transformed potential, $\hat \phi(\vec{p},z)$,
\begin{eqnarray}
 \label{Eq:poissonz>0}
 -\frac{d^2\hat \phi}{dz^2}+p^2\hat \phi(\vec{p},z)=4\pi q\delta(z-d) &&\quad\quad (z>0), \\
 %
  \label{Eq:poissonz<0}
 -\epsilon\frac{d^2\hat \phi}{dz^2}+\epsilon p^2\hat \phi(\vec{p},z)=0 &&\quad\quad (z<0),
\end{eqnarray}
The equations for the electric potential Eqs.~(\ref{Eq:poissonz>0},\ref{Eq:poissonz<0}) have to obey the four boundary conditions
\begin{enumerate}
	\item $\hat \phi(\vec{p},z=-\eta)=\hat \phi(\vec{p},z=+\eta)$
	
	\item $\left.\frac{d\hat \phi}{dz}\right|_{z=+\eta}-\epsilon\left.\frac{d\hat \phi}{dz}\right|_{z=-\eta}=\kappa
	|\vec{p}|\hat \phi(\vec{p},z=0) $
	
	\item $\hat \phi(\vec{p},z=d-\eta)=\hat \phi(\vec{p},z=d+\eta)$
	
	\item $\left.\frac{d\hat \phi}{dz}\right|_{z=d-\eta}-\left.\frac{d\hat \phi}{dz}\right|_{z=d+\eta}=4\pi q $
\end{enumerate}
where $\kappa=(1/2)(\alpha\theta/\pi)^2$. 
The boundary condition 2 above follows directly by inserting Eq. (\ref{Eq:B}) into the Poisson equations, Eqs. (\ref{Eq:Poisson-1}) and 
(\ref{Eq:Poisson-2}) 
and performing a Fourier transform in the plane.  
These boundary conditions are used to determine the unknown coefficients by matching the solutions in three regions,
\begin{eqnarray}
\hat \phi(\vec{p},z)=Ae^{|\vec{p}|z}   &&  \quad\quad (z<0), \\
%
\hat \phi(\vec{p},z)=Be^{|\vec{p}|z}+Ce^{-|\vec{p}|z} && \quad\quad (0<z<d), \\
%
\hat \phi(\vec{p},z)=De^{-|\vec{p}|z} && \quad\quad (z>d).
\end{eqnarray}
After determining $A$, $B$, $C$, and $D$, we obtain, 
\begin{equation}
\hat{\phi}(\vec{p},z>0)=\frac{2\pi q}{p}\left[\left(\frac{ 1-\epsilon-\kappa}{1+\epsilon+
	\kappa}\right)e^{-p(z+d)}+e^{-p|z-d|}
\right],
\end{equation}
\begin{equation}
\hat{\phi}(\vec{p},z<0)=\frac{4\pi q}{1+\epsilon+
	\kappa}\frac{e^{-p|z-d|}}{p}.
\end{equation}
Since, 
\begin{equation}
2\pi \int\frac{d^2p}{(2\pi)^2}\frac{e^{i\vec{p}\cdot\vec{r}-p|z-z_0|}}{p}
=\frac{1}{\sqrt{r^2+(z-z_0)^2}},
\end{equation}
where $z_0\in\mathbb{R}$, 
we easily obtain the electric potential, 
\begin{eqnarray}
\label{Eq:potentialz>0}
\phi(r,z>0)&=&q\left[\left(\frac{ 1-\epsilon-\kappa}{1+\epsilon+
	\kappa}
\right)\frac{1}{\sqrt{r^2+(z+d)^2}}\right.\nonumber\\
&+&\left.\frac{1}{\sqrt{r^2+(z-d)^2}}\right], \\
%
\label{Eq:potentialz<0}
\phi(r,z<0)&=&\frac{2q}{ 1+\epsilon+\kappa}\frac{1}{\sqrt{r^2+(z-d)^2}},
\end{eqnarray}
yielding in turn the   
electric fields for $z<0$ and $z>0$, 
\begin{eqnarray}
\label{Eq:E-z+}
\vec{E}(\vec{r},z>0)&=&q\left[\left(\frac{ 1-\epsilon-\kappa}{1+\epsilon+
	\kappa}
\right)\frac{\vec{r}+(z+d)\hat{\vec{z}}}{[r^2+(z+d)^2]^{3/2}}
\right.\nonumber\\
&+&\left.\frac{\vec{r}+(z-d)\hat{\vec{z}}}{[r^2+(z-d)^2]^{3/2}}\right],
\end{eqnarray}
\begin{equation}
\label{Eq:E-z-}
\vec{E}(\vec{r},z<0)=\frac{2q}{ 1+\epsilon+\kappa}\frac{\vec{r}+(z-d)\hat{\vec{z}}}{[r^2+(z-d)^2]^{3/2}}.
\end{equation}
Unremarkably, the above expressions reduce to the standard textbook ones when $\theta=0$. Note that only $\vec{E}(\vec{r},z=0)$ is needed to determine the magnetic 
field via Eq. (\ref{Eq:B}). In view of the axion term, the electric field is discontinuous at $z=0$, as evidenced by the boundary conditions above. 
Thus, we have, 
\begin{eqnarray}
\label{Eq:Ez+}
\vec{E}(\vec{r},z=+\eta)=\frac{2q}{[ 1+\epsilon+\kappa](r^2+d^2)^{3/2}}  \times
\\
\left\{\vec{r} -d\left[\epsilon-\kappa\right]\hat{\vec{z}}\right\},
\end{eqnarray}
\begin{equation}
\vec{E}(\vec{r},z=-\eta)=\frac{2q}{ 1+\epsilon+\kappa}\frac{\vec{r}-d\hat{\vec{z}}}{(r^2+d^2)^{3/2}}.
\end{equation}
Therefore, 
\begin{equation}
\hat{\vec{z}}\times\vec{E}(\vec{r},z=0)=\frac{2q}{ 1+\epsilon+\kappa}\frac{\hat{\vec{z}}\times\vec{r}}{(r^2+d^2)^{3/2}}.
\end{equation}

 \subsection{Calculation of the magnetic field}
 The  most straightforward 
 way to calculate the magnetic field is by inserting the expression for the electric field directly in Eq. (\ref{Eq:A}) and 
 performing the resulting integral. The calculation is considerably easier using a Fourier transform on the TI surface, in which 
 case we obtain from Eq. (\ref{Eq:A}),
\begin{eqnarray}
\hat{\vec{A}}(\vec{p},z)&=&\frac{\alpha\theta}{2\pi}\frac{e^{-p|z|}}{p}\hat{\vec{z}}\times\hat{vec{E}}(\vec{p},z=0)
\nonumber\\
&=&-N\Phi e^{-p(|z|+d)}\frac{\hat{\vec{z}}\times i\vec{p}}{p^2},
\end{eqnarray}
where we have assumed $q=Ne$, $N\in\mathbb{Z}$, and, 
\begin{equation}
\Phi=\frac{\alpha^2\theta\Phi_0}{\pi(1+\epsilon+\kappa)},
\label{Eq:flux}
\end{equation}
with $\Phi_0=hc/e$ being the elementary flux quantum.  
Thus, we obtain the Fourier representation for the vector potential, 
\begin{equation}
\vec{A}(\vec{r},z)=-N\Phi(\hat{\vec{z}}\times\nablab a(\vec{r},z)),
\end{equation}
where,
\begin{equation}
a(\vec{r},z)=\int\frac{d^2p}{(2\pi)^2}\frac{e^{-p(|z|+d)+i\vec{p}\cdot\vec{r}}}{p^2}.
\end{equation}
Performing the angular integration in the above equation yields, 
\begin{equation}
a(\vec{r},z)=\frac{1}{2\pi}\int_{0}^{\infty}\frac{dp}{p}e^{-p(|z|+d)}J_0(pr),
\end{equation}
where $J_0(x)$ is a Bessel function. Thus, 
\begin{eqnarray}
\nablab a&=&-\frac{\vec{r}}{2\pi r}\int_0^\infty dp J_1(pr)e^{-p(|z|+d)}
\nonumber\\
&=&-\frac{\vec{r}}{2\pi r^2}\left[1-\frac{(|z|+d)}{\sqrt{r^2+(|z|+d)^2}}\right].
\end{eqnarray}

 Thus, by performing the inverse Fourier transform, we obtain, 
 \begin{equation}
 \label{Eq:A-vortex}
 \vec{A}(\vec{r},z)=\frac{N\Phi}{2\pi}\frac{\hat{\vec{z}}\times\vec{r}}{r^2}
 \left[1-\frac{(|z|+d)}{\sqrt{r^2+(|z|+d)^2}}\right],
 \end{equation}
 whose curl yields, 
 \begin{equation}
 \label{Eq:B-vortex}
 \vec{B}(\vec{r},z)=\frac{N\Phi}{2\pi}\sum_{s=\pm}sH(sz)\frac{\vec{r}+(z+sd)\hat{\vec{z}}}{[r^2+(z+sd)^2]^{3/2}},
 \end{equation}
 where $H$ is the Heaviside step function. It immediately follows that,
 \begin{eqnarray}
 \label{Eq:divless}
 \nablab\cdot\vec{B}&=&\frac{N\Phi}{2\pi}\delta^2(\vec{r})[H(z)\delta(z+d)-H(-z)\delta(z-d)]
 \nonumber\\
 &=&\frac{N\Phi}{2\pi}\delta^2(\vec{r})[H(-d)\delta(z+d)-H(-d)\delta(z-d)]
 \nonumber\\
 &=&0,
 \end{eqnarray}
 since $H(-d)=0$.  Although the total magnetic flux through any closed surface containing a sphere of radius $d$ centered at the origin 
 vanishes,  we note that the flux through the TI surface yields precisely $\Phi$ for all z.
 
 \begin{figure}
	\centering
	\includegraphics[width=1.0\columnwidth]{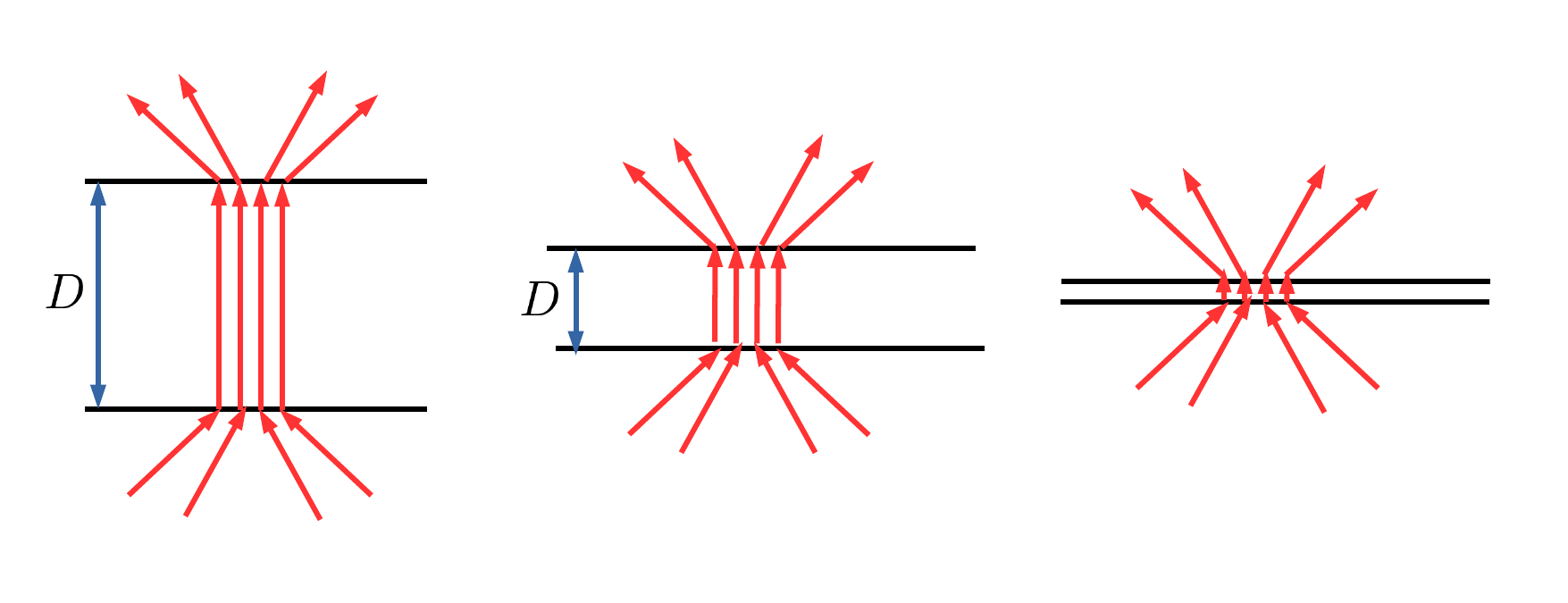}
	\caption{(Color online) Schematic depiction of a Pearl vortex \cite{Pearl-1964} as the small thickness limit of an Abrikosov-Nielsen-Olesen vortex 
		\cite{A-vortices,NO-vortices} inside a 
		superconducting slab of thickness $D$. As $D\to 0$, the vortex line approaches a point vortex in a very thin superconducting thin film.   
	}
	\label{Fig:Pearl-Vortex}
\end{figure}
For $d\to 0$ we obtain from Eq. (\ref{Eq:A-vortex}) that 
 \begin{equation}
 \label{Eq:A0}
 \vec{A}(\vec{r},z)=\frac{N\Phi}{2\pi}\frac{\hat{\vec{z}}\times\vec{r}}{r^2}
 \left(1-\frac{|z|}{\sqrt{r^2+z^2}}\right).
 \end{equation}
 We see that if $|z|$ were replaced by $z$ in Eq. (\ref{Eq:A0}), it would precisely yield the vector potential of a straight vortex line  (or Dirac string) over the negative $z$-axis ending at a magnetic monopole at $z=0$. This fact is crucial and it is what makes Eq. (\ref{Eq:A0}) to correspond to the magnetic field 
 of a Pearl vortex  
\begin{equation}
\label{Eq:Pearl-B}
\vec{B}(r,z)=\frac{N\Phi}{2\pi}{\rm sgn}(z)\frac{\vec{r}+z\hat{\vec{z}}}{(r^2+z^2)^{3/2}},
\end{equation}
where for a superconductor $\Phi=n\Phi_0/2$. 
An important difference between the magnetic field above and the actual Pearl vortex arising in superconductors is that the 
former holds for all $r$, no matter small, while the actual Pearl vortex field profile of Eq. (\ref{Eq:Pearl-vortex}) follows from the large distance limit $r\gg \lambda_L$ for a flux line of vanishing length. In other words, Eq. (\ref{Eq:Pearl-B}) describes an {\it exact Pearl} vortex.

By removing the ${\rm sgn}(z)$ factor in Eq. (\ref{Eq:Pearl-B}) we obtain precisely the magnetic field of a Dirac magnetic monopole: in other words the magnetic field (\ref{Eq:Pearl-B}) 
behaves as a monopole for $z>0$ and as an anti-monopole for $z<0$, yielding in this way $\nablab\cdot\vec{B}=0$, see Fig.\ref{Fig:Pearl-Vortex}. 
 
\begin{figure}
	\includegraphics[width=1.0\columnwidth]{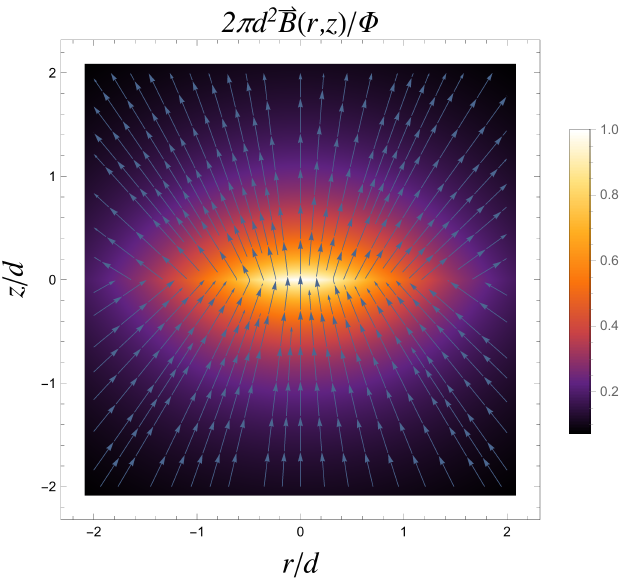}
	\caption{Stream density plot of the magnetic field for an electric charge $q$ at a $d\neq 0$ above the surface of a (topological) magnetoelectric.}
	\label{Fig:Magnetic-field}
\end{figure}

The stream density plot associated to the magnetic field components above is shown in Fig. \ref{Fig:Magnetic-field} for the reduced coordinates $z/d$ and $r/d$. We note the presence of an 
extended solitonic object near $z=0$, indicating that the point vortex 
becomes for $d\neq 0$ a kind of pancake vortex. 
 
The magnetic field (\ref{Eq:B-vortex}) can ibe interpreted as corresponding to image magnetic charges 
of strength $g_\pm=\pm N\Phi/(2\pi)$ located at $z_\pm=\mp d$, respectively. Thus., the magnetic charge $g_+$ at $z=-d$ mirrors the 
magnetic field at $z>0$, while the magnetic charge $g_-$ at $z=d$ mirrors the magnetic field at $z<0$.  However, we have seen in 
Eq. (\ref{Eq:divless}) that $\nablab\cdot\vec{B}$ vanishes everywhere. 
%

\section{Angular momentum} 

\subsection{Electromagnetic field contribution to the angular momentum}

The electromagnetic momentum density $\Pib=(\vec{E}\times\vec{B})/(4\pi c)$ is nonzero for all $z>0$, vanishing 
otherwise. Thus, the angular momentum carried by the electromagnetic field is obtained by integrating over 
all space the angular momentum density, $(\vec{r}+z\hat{\vec{z}})\times\Pib$. 
The components $L_x$ and $L_y$ vanish due rotational invariance in the plane, so we obtain, 
\begin{eqnarray}
&&L_z=-\frac{N^2e\Phi d}{2\pi c}
\nonumber\\
&\times&\int_0^\infty dz\int_0^\infty \frac{dr~r^3}{[r^2+(z-d)^2]^{3/2}[r^2+(z+d)^2]^{3/2}}.
\nonumber\\
\end{eqnarray}
Performing the integral above yields, 
\begin{equation}
\label{Eq:Lz}
L_z=-N^2\frac{\Phi}{\Phi_0}\frac{\hbar}{2}.
\end{equation} 
As with angular momentum of dyons \cite{Schwinger-1969}, the above angular momentum is independent of the distance 
of the point charge to the TI surface. 
However, in contrast to the dyon, the above angular momentum is neither an integer nor 
a half-integer multiple of $\hbar$. In order for this to happen, $\Phi=n\Phi_0$, $n\in\mathbb{Z}$, would have to be satisfied. 
But this would then in general imply a negative dielectric constant for all $n$ as according to Eq.~(\ref{Eq:flux}).
Interestingly however, $L_z$ features a quantization $\sim N^2$ characteristic of angular momentum of Chern-Simons vortices \cite{Jackiw-Weinberg}. {\color{black} Despite this similarity, it is important to emphasize that the integer squared arising in the latter corresponds to the vortex quantum number rather than the number of particles $N$ as used here in connection with the point charge $q=Ne$. This fact has experimental consequences and will come back to this point below.} 
 
It might be tempting to associate the result of Eq. (\ref{Eq:Lz}) to a dyon exhibiting anyon behavior~\cite{Qi_image_monopole}. As
$\Phi \propto {(1+\epsilon+\kappa)^{-1}}$ with $\epsilon$ the dielectric constant of the magnetoelectric medium and $\kappa \propto \alpha^2$ this would imply a dependence of $L_z$ on  the dielectric and fine structure constant.
Such non-universal behavior in the process of exchanging two particles is rather not expected in quantum statistics. 

{\color{black} Let us compare the result of Eq. (\ref{Eq:Lz}) with the one obtained for a vortex at the interface of a SC-TI heterostructure. For the latter it has been demonstrated that \cite{Nogueira-van_den_Brink-Nussinov_London_axion}, 
\begin{equation}
	\label{Eq:L-z-SC-TI}
	L_z^{\rm SC-TI}=-\frac{n^2\theta}{4\pi}\frac{\hbar}{2}, 
\end{equation}
where $n$ is the vortex quantum number. The integer number square appearing in both  Eqs. (\ref{Eq:Lz}) and (\ref{Eq:L-z-SC-TI}) have a completely different origin. In Eq.  (\ref{Eq:Lz}) $N$ refers to the number of electron charge units contained in the charge $q$ at a distance $d$ from the TI surface, while $n$ in Eq. (\ref{Eq:L-z-SC-TI}) is a winding number associated to flux quantization in superconductors. It turns out that $n$ is rarely larger than the unity. On the other hand, it is easily possible to have considerably larger values of $N$. Thus, in spite of the small factor $\alpha^2$ in Eq. (\ref{Eq:L-z-SC-TI}), $L_z$ is not necessarily small and it can actually be of the same order or even larger than the induced angular momentum in SC-TI systems.

Another important difference between Eq. (\ref{Eq:L-z-SC-TI}) and Eq. (\ref{Eq:Lz}) is that the former refers to the {\it total} angular momentum rather than just the one due to the electromagnetic field. Equation (\ref{Eq:L-z-SC-TI}) actually contains also the mechanical contribution, i.e., one corresponding to the orbital motion of Cooper pairs around the vortex. In the next subsection we will address the contribution of the surface fermions to the total angular momentum. 

Interestingly, conservation of angular momentum would imply that the induced magnetoelectric angular momentum be counterbalanced by a mechanical rotation of the TI. This can be in principle be detected experimentally. 
}  

\subsection{Angular momentum contribution from Surface Dirac fermions}
\label{Sec:zero-modes}

The Dirac equation on the TI surface is given by, 
\begin{equation}
	H\psi_E=\frac{E}{v_F}\psi_E,
\end{equation}
with the Hamiltonian,
\begin{equation}
H=\alphab\cdot\left[-i\hbar\nablab-\frac{e}{c}\vec{A}(\vec{r},0)\right]+\frac{\Delta}{v_F}+\frac{e}{v_F}\phi(r,0),
\end{equation}
where $v_F$ is the Fermi velocity and in terms of Pauli matrices $\alphab=(-\sigma_y,\sigma_x)$. The Pauli matrix $\sigma_z$ multiplies $\Delta=-\mu_BL_z$ ($\mu_B$ is the Bohr magneton), which is a Zeeman term induced by the angular momentum of the electromagnetic field. Were we to have $\phi=0$ and $\Delta=0$, the state $\sigma_z\psi_E$ would have been 
an eigenstate of the Dirac operator with energy $-E$, and therefore $\sigma_z\psi_E=\psi_{-E}$ \cite{Jackiw-1984}. This property fails when 
$\phi\neq 0$, which is precisely the case we are dealing with here. This implies that any underlying zero modes in this problem are not self-conjugate. 


{\color{black} There are two distinct situations to be considered here, depending on whether $d=0$ or not. From Eq. (\ref{Eq:A-vortex}), we have, 
\begin{equation}
\vec{A}(\vec{r},0)=\frac{N\Phi}{2\pi}\frac{\hat{\varphib}}{r}\left(1-\frac{d}{\sqrt{r^2+d^2}}\right),
\end{equation}
so we see that for $d\neq 0$ the vector potential vanishes for $r\to 0$ and behaves as $\vec{A}(\vec{r},0)\sim \hat{\varphib}/r$ for $r\gg d$. For $d=0$, on the other hand, the potential is singular at $r=0$, but the Dirac equation can still be solved by employing a singular gauge transformation. 

Let us consider the case where the charge lies exactly on the TI surface, so that we can set $d=0$. In this case, performing the unitary transformation $U=e^{iN(\Phi/\Phi_0)\varphi}$ casts the Hamiltonian in the form, 
\begin{equation}
	U^\dagger H U=-i\hbar\alphab\cdot\nablab+\frac{\Delta}{v_F}\sigma_z+\frac{e}{v_F}\phi(r,0),
\end{equation}
and induces a multivalued phase factor in the  spinor field,  $\psi_E=e^{iN(\Phi/\Phi_0)\varphi}\Psi_E$. Hence, the Dirac equation becomes, 
 \begin{eqnarray}
 	\label{Eq:Gauged-Dirac}
&\left(-i\alphab\cdot\nablab+m\sigma_z+\frac{Nb\Phi}{\Phi_0r}\right)\Psi_E=\frac{\omega}{v_F}\Psi_E,
 \end{eqnarray}
 where $b=2\pi c/(\alpha\theta v_F)$, $m=\Delta/(\hbar v_F)$, and $\omega=E/\hbar$. 
Solutions of the Dirac equation above have the form, 
 \begin{equation}
 	\label{Eq:psi0}
 	\Psi_E=\left[
 	\begin{array}{c}
 		u \\
 		\noalign{\medskip}
 		v
 	\end{array}
 	\right]
 	=\left[
 	\begin{array}{c}
 		e^{i(n-1/2)\varphi}f(r) \\
 		\noalign{\medskip}
 		e^{i(n+1/2)\varphi}g(r)
 	\end{array}
 	\right],
 \end{equation}
 where $n$ is an integer. Note that $\Psi_E$ satisfies anti-periodic boundary conditions. It follows that $\psi_E$ is an eigenstate of the angular momentum operator, 
 \begin{eqnarray}
 	J_z=-i\hbar\frac{\partial}{\partial\varphi}+\frac{\hbar}{2}\sigma_z,
 \end{eqnarray}
 and thus, 
 \begin{equation}
 	\label{Eq:Jz}
 	J_z\psi_E=\hbar\left(n+\frac{N\Phi}{\Phi_0}\right)\psi_E.
 \end{equation}

The total angular momentum of the system is given by adding the above result to the electromagnetic field contribution of Eq. (\ref{Eq:Lz}), 
 \begin{equation}
 	J_z^{\rm tot}=\hbar \left[n-N\left(\frac{N}{2}-1\right)\frac{\Phi}{\Phi_0}\right]. 
 \end{equation}
 
 The Ansatz (\ref{Eq:psi0}) leads to the pair of equations, 
 %
 %
 \begin{equation}
	\label{Eq:g-1}
	\frac{dg}{d r}+\frac{1}{r}\left(n+\frac{1}{2}\right)g+\frac{\beta}{r}f=(\Omega-m)f,
\end{equation}
\begin{equation}
	\label{Eq:f-1}
	\frac{d f}{d r}-\frac{1}{r}\left(n-\frac{1}{2}\right)f-\frac{\beta}{r}g=-(\Omega+m)g,
\end{equation}
where we have defined $\Omega=\omega/v_F$ and 
$\beta=Nb\Phi/\Phi_0=2N(c/v_F)\alpha/(1+\epsilon+\kappa)$. 
%
Because of the way the equations are coupled, it is more convenient to find solutions by introducing the new variables $F=f+g$ and $G=f-g$, in which case the equations are recast in the form, 
\begin{equation}
	\label{Eq:F}
	\frac{dF}{dr}+\frac{F}{2r}+\frac{1}{r}(\beta-n)G=-mF+\Omega G,
\end{equation}
\begin{equation}
	\label{Eq:G}
	\frac{dG}{dr}+\frac{G}{2r}-\frac{1}{r}(\beta+n)F=-mG-\Omega F.
\end{equation}

Normalized zero mode solutions ($\Omega=0$) are obtained in the form $F(r)=F_0e^{-mr}r^s$ and $G(r)=G_0e^{-mr}r^s$, assuming $m>0$. These normalized solutions are obtained for 
$s=\sqrt{n^2-\beta^2}-1/2$.
The boundary conditions at $r=0$ require that 
\begin{equation}
	\label{Eq:ineq}
	n^2>\beta^2+\frac{1}{4}.
\end{equation}

The constants $f_0=(F_0+G_0)/2$ and $g_0=(F_0-G_0)/2$ are easily determined from the normalization condition to be given by, 
\begin{equation}
	f_0^2=\frac{(n-\beta)(2m)^{1+2\sqrt{n^2-\beta^2}}}{2\pi n\Gamma(1+2\sqrt{n^2-\beta})},
\end{equation}
\begin{equation}
	g_0^2=\frac{\beta(2m)^{1+2\sqrt{n^2-\beta^2}}}{2\pi n\Gamma(1+2\sqrt{n^2-\beta})}.
\end{equation}

}

Finally, let us comment on an aspect of zero modes that is usually true in most topological systems but that does not hold in the present case, namely, 
the existence of an index theorem \cite{Weinberg_PhysRevD.24.2669} or, more specifically, the Atiyah-Patodi-Singer (APS) index theorem 
\cite{atiyah_patodi_singer_1975,nakahara2003geometry}. This theorem states that the number of positive energy modes minus the negative 
energy ones is an integer topological invariant, the so called $\eta$-invariant. In order for this theorem be applicable it is necessary to be able to 
map one-to-one the positive energy modes to the negative ones, with the zero modes being self-conjugate. When $\phi=0$ this mapping is 
provided by the Pauli matrix $\sigma_z$, similarly to the situations encountered in Refs. \cite{Jackiw-1984} and \cite{Jackiw-Rossi} 
(see also Ref. \cite{Weinberg_PhysRevD.24.2669}), where the Pauli matrix $\sigma_z$ acts as conjugation matrix. As already mentioned, in our case 
$\sigma_z$ does not map $\psi_E$ into $\psi_{-E}$, so we are unable to apply the  APS index theorem in this case.

 \section{General argument for absence of induced magnetic monopoles}
\label{Sec:General}
The constraint $\nablab\cdot\vec{B}=0$ not being violated in the general axion magnetoelectric screening problem can also be argued on the basis of more general considerations. Due to the $\nablab\cdot\vec{B}=0$ constraint a string singularity has to be  attached to a Dirac monopole. Monopoles without strings are only possible if topologically nontrivial gauge transformations are allowed \cite{Wu-Yang-PhysRevD.12.3845}, in which case two nonsingular vector potentials can be used, $\vec{A}_\pm(\vec{r},z)=\pm gr^{-2}(\hat{\vec{z}}\times\vec{r})(1\mp z/\sqrt{r^2+z^2})$, defined in the regions of a sphere around a point monopole $g$ excluding the south and north poles, respectively.  These gauge potentials differ by a singular gauge transformation, since $\vec{A}_+-\vec{A}_-=2g\nablab\varphi$, $\varphi\in[0,2\pi]$. Indeed, $\nablab\times\nablab\varphi=2\pi\delta^2(\vec{r})\hat{\vec{z}}$. As a consequence, $\nablab\cdot\vec{B}=4\pi g\delta^2(\vec{r})\delta(z)$. The corresponding topologically nontrivial gauge transformation is therefore ${\cal G}=\exp[2ieg\varphi/(\hbar c)]$, which leads to the Dirac condition, $eg/(\hbar c)=n/2$, $n\in\mathbb{Z}$. Such a scenario is not realizable within the axion electrodynamics discussed here, where gauge transformations are topologically trivial. In this case magnetic monopoles would require non-vanishing currents at large distances, contradicting one of the basic tenets of electromagnetism.  

 \section{Conclusion} 
 We have obtained the induced magnetic field due to a charged particle above the surface of a topological insulator or any magnetoelectric material in general. The exact magnetic field has been obtained directly without using image magnetic charges. The solution allows nevertheless for a clear identification of  the reflection and transmission image magnetic charges. However, the latter cannot be interpreted as induced magnetic  monopoles, since the vector potential does not allow for topologically nontrivial gauge transformations and there is no flux tube connecting the magnetic image charges.  In the limit case where the charge lies exactly at the surface, the field of a  point vortex, also known as Pearl vortex \cite{Pearl-1964} is obtained. Such a point vortex resembles a monopole, but it is quite different  from it, as it does not allow for topologically nontrivial gauge transformations necessary to make $\nablab\cdot\vec{B}\neq 0$. 
 
 {\color{black} We have found that the vortex-like solution features a nontrivial angular momentum for the electromagnetic field, in a situation reminiscent to the vortex solution for a TI proximate to a type II superconductor obtained in Ref. \cite{Nogueira-van_den_Brink-Nussinov_London_axion}. As in the latter reference, we calculated the angular momentum exactly. However, there is an essential difference between them, namely, the result we have obtained in Eq. (\ref{Eq:Lz}) has a dependence on the dielectric constant, while such a dependence cancels out in the calculation leading to Eq. (\ref{Eq:L-z-SC-TI}), corresponding to the results obtained in Ref. \cite{Nogueira-van_den_Brink-Nussinov_London_axion}. The lack of dependence on the dielectric constant in Eq. (\ref{Eq:L-z-SC-TI}) is due to the fact that this angular momentum corresponds to both mechanical and electromagnetic contribution. However, accounting for the mechanical contribution from the surface Dirac fermions in the presence of an external point charge does remove the $\epsilon$-dependence in the total angular momentum. For this reason, one cannot claim that the system exhibits fractional statistics, since quantum statistics should be a universal property of the system and not be dependent on specific material properties like the dielectric constant. Nevertheless the result of an induced total angular momentum is a very interesting one, since it indicates that a point charge in proximity to a TI would induce a compensating mechanical torque that can be detected experimentally.}
 
 Since the Dirac equation is known to favor zero mode solutions in the presence of topological defects like vortices or monopoles \cite{Jackiw-Rossi}, 
 we have also investigated this possibility here and obtained that zero modes indeed exist. However, the fermions have to be gapped and thus break TR, otherwise no normalized zero modes can be found. 

\acknowledgments

JvdB acknowledges support from the Deutsche Forschungsgemeinschaft (DFG) through the W\"urzburg- Dresden Cluster of Excellence on Complexity and Topology in Quantum Matter—ct.qmat (EXC 2147, Project No. 39085490) and the Collaborative Research Center (Sonderforschungsbereich) SFB 1143 (Project No. 247310070).

\bibliography{screening}

\end{document}